\documentclass[prl,twocolumn]{revtex4}
 \usepackage[english]{babel}
 \usepackage{epsfig,amssymb,amsmath}
 \usepackage{color}

\def\work{Letter}

\begin{document}

\title{Nonequilibrium excitations of molecular vibrons}

\author{D.\,A.~Ryndyk\footnote{On leave from the Institute for Physics of Microstructures, RAS,
 Nizhny Novgorod, Russia}, M.~Hartung, and G.~Cuniberti}

\affiliation{Institut f\"ur Theoretische Physik, Universit\"at Regensburg, D-93040 Germany}

\begin{abstract}

\end{abstract}

\begin{abstract}
We consider the nonequilibrium quantum vibrations of a molecule clamped between two
macroscopic leads in a current-carrying state at finite voltages. Our approach is
based on the nonequilibrium Green function technique and the self-consistent Born
approximation. Kinetic equations for the average populations of electrons and vibrons
are formulated in the weak electron-vibron coupling case and self-consistent solutions
are obtained. The effects of vibron emission and vibronic instability are demonstrated
using few-orbital models. The importance of the electron-vibron resonance is shown.
\end{abstract}

\date{\today}
\maketitle

During the past several years, nonequilibrium quantum transport in nanostructures and,
in particular, transport through single molecules, has been in the focus of both
experimental and theoretical investigations because of possible electronic device
applications. Recently, the interaction of electrons with molecular vibrations
attracted attention after experiments on inelastic electron transport through single
molecules
\cite{Cuniberti05springer,Reed97science,Park00nature,Park02nature,Liang02nature,
Smith02nature,Zhitenev02prl,Yu}. New theoretical treatments were presented in
Refs.\,\cite{Lundin02prb,Zhu03prb,Braig03prb,Aji03condmat,Mitra04prb,Frederiksen04master,
Frederiksen04prl,MTU,GRN,Ryndyk05prb,Hartung2004,Cizek05prb,Koch}. In this \work , we
consider a quantum theory of nonequilibrium vibronic excitation.

Basically there are two main nonequilibrium effects: the electronic spectrum
modification \cite{Ryndyk05prb} and excitation of vibrons (quantum vibrations). In the
weak electron-vibron coupling case the spectrum modification is usually small (which
is dependent, however, on the vibron dissipation rate, temperature, etc.) and the main
possible nonequilibrium effect is the excitation of vibrons at finite voltages. We
develop an analytical theory for this case. This theory is based on the
self-consistent Born approximation (SCBA), which allows to take easily into account
and calculate nonequilibrium distribution functions of electrons and vibrons.

If the mechanical degrees of freedom are coupled strongly to the environment
(dissipative vibron), then the dissipation of molecular vibrations is determined by
the environment. However, if the coupling of vibrations to the leads is weak, we
should consider the case when the vibrations are excited by the current flowing
through a molecule, and the dissipation of vibrations is also determined essentially
by the coupling to the electrons. In this \work , we show that the effects of vibron
emission and vibronic instability are important especially in the case of
electron-vibron resonance.

We describe a molecule coupled to free conduction electrons in the leads by a usual
tunneling Hamiltonian. Furthermore, the electrons are coupled to vibrational modes. We
do not consider Coulomb interaction to avoid further effects, such as Coulomb blockade
and Kondo effect, which could dominate over the physics which we want to address,
however self-consistent mean-field effects can be included easily in our approach. The
full Hamiltonian is the sum of the molecular Hamiltonian $\hat H_M$, the Hamiltonians
of the leads $\hat H_{R(L)}$, the tunneling Hamiltonian $\hat H_T$ describing
molecule-lead coupling, the vibron Hamiltonian $\hat H_V$ including electron-vibron
coupling and coupling of vibrations to the environment
\begin{equation}\label{H}
 \hat H=\hat H_M+\hat H_L+\hat H_R+\hat H_T+\hat H_V.
\end{equation}

A molecule (as well as a system of small quantum dots) is described by a set of
localized states $|\alpha\rangle$ with energies $\epsilon_\alpha$ (tight-binding
model) by the following model Hamiltonian:
\begin{equation}\label{H_M}
 \hat H^{(0)}_M=\sum_\alpha\left(\epsilon_\alpha+e\varphi_\alpha(t)\right)
 d^{\dag}_\alpha d_\alpha +
 \sum_{\alpha\neq\beta}t_{\alpha\beta}
 d^{\dag}_\alpha d_\beta,
\end{equation}
where $d^{\dag}_\alpha$,$d_\alpha$ are creation and annihilation operators in the
states $|\alpha\rangle$, and $\varphi_\alpha(t)$ is the (self-consistent) electrical
potential.

\begin{figure}[b]
\begin{center}
\epsfxsize=0.7\hsize
\epsfbox{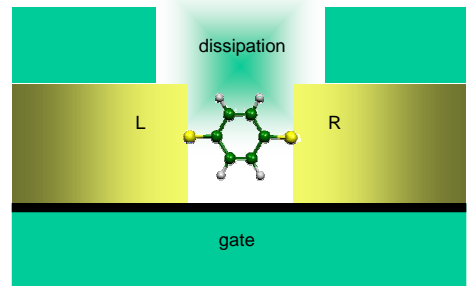}
\caption{(Color online) Schematic picture of the considered system.}
\label{fig1}
\end{center}
\end{figure}

The Hamiltonians of the right (R) and left (L) leads are
\begin{equation}
 \hat H_{i=L(R)}=\sum_{k\sigma}(\epsilon_{ik\sigma}+e\varphi_i(t))
 c^{\dag}_{ik\sigma}c_{ik\sigma},
\end{equation}
$\varphi_i(t)$ are the electrical potentials of the leads, and the tunneling
Hamiltonian
\begin{equation}\label{H_T}
 \hat H_T=\sum_{i=L,R}\sum_{k\sigma,\alpha}\left(V_{ik\sigma,\alpha}
 c^{\dag}_{ik\sigma}d_\alpha+h.c.\right)
\end{equation}
describes hopping between the leads and the molecule. Direct hopping between two leads
is neglected (weak molecule-lead coupling case).

Vibrations and the electron-vibron coupling are described by the Hamiltonian
\cite{Frederiksen04master,Frederiksen04prl,Hartung2004}
\begin{equation}\label{H_V}
 \hat H_{V}=\sum_q\hbar\omega_qa_q^\dag a_q
 +\sum_{\alpha\beta}\sum_qM^q_{\alpha\beta}(a_q+a_q^\dag)
 d^{\dag}_\alpha d_\beta.
\end{equation}
Here vibrations are considered as localized phonons and $q$ is an index labelling
them, not the wave-vector. We include both diagonal coupling, which describes a change
of the electrostatic energy with the distance between atoms, and the off-diagonal
coupling, which describes the dependence of the matrix elements $t_{\alpha\beta}$ over
the distance between atoms.

We use the nonequilibrium Green function (NGF) method \cite{Kadanoff62book,Keldysh64},
which now is a standard approach in mesoscopic physics and molecular electronics
\cite{Cuniberti05springer}. We follow the formulation of Meir, Wingreen, and Jauho
\cite{Meir92prl,Jauho94prb,Haug96book}, which has been already applied to the case of
self-consistency in Ref.~\cite{Ryndyk05prb}.

The current in the left ($i=L$) or right ($i=R$) contact to the molecule is described
by the well-known expression
\begin{equation}\label{J}\begin{array}{c}\displaystyle
 J_{i=L,R}=\frac{ie}{\hbar}\int\frac{d\epsilon}{2\pi}{\rm Tr}\left\{
 {\bf\Gamma}_i(\epsilon-e\varphi_i)\left({\bf G}^<(\epsilon)+ \right.\right.\\[0.5cm]
 \displaystyle \left.\left.
 +f^0_i(\epsilon-e\varphi_i)
 \left[{\bf G}^R(\epsilon)-{\bf G}^A(\epsilon)\right]\right)\right\},
 \end{array}
\end{equation}
where $f^0_i(\epsilon)$ is the equilibrium Fermi distribution function with chemical
potential $\mu_i$, and the level-width function is
%
$${\bf\Gamma}_{i=L(R)}(\epsilon)=
\Gamma_{i\alpha\beta}(\epsilon) =2\pi\sum_{k\sigma}
V_{ik\sigma,\beta}V^*_{ik\sigma,\alpha}\delta(\epsilon-\epsilon_{ik\sigma}).$$

The matrix lesser (retarded, advanced) Green functions of a nonequilibrium molecule
${\bf G}^{<(R,A)}\equiv G_{\alpha\beta}^{<(R,A)}$ can be found from the Dyson-Keldysh
equations in the integral form or from the corresponding equations in the differential
form \cite{Ryndyk05prb} (and references therein).

In the standard self-consistent Born approximation, using the Keldysh technique, one
obtains for the vibronic self-energies
\cite{Rammer86rmp,Haug96book,MTU,Mitra04prb,GRN,Hartung2004,Frederiksen04master,Frederiksen04prl}
\begin{eqnarray}\label{SigmaRA}
& \displaystyle {\bf\Sigma}^{R(V)}(\epsilon)=\frac{i}{2}\sum_q\int\frac{d\omega}{2\pi}
 \left({\bf M}^q{\bf G}^{R}_{\epsilon-\omega}{\bf M}^qD^K_{q\omega}+
 \right. \nonumber \\[0.2cm]
& \displaystyle \left. +{\bf M}^q{\bf G}^K_{\epsilon-\omega}{\bf M}^qD^{R}_{q\omega}
 -2D^{R}_{q\omega=0}
 {\bf M}^q{\rm Tr}\left[{\bf G}^<_{\omega}{\bf M}^q\right]\right), \\[0.3cm]
 \label{SigmaK}
& \displaystyle {\bf\Sigma}^{<(V)}(\epsilon)=i\sum_q\int\frac{d\omega}{2\pi}{\bf M}^q
 {\bf G}^<_{\epsilon-\omega}{\bf M}^qD^<_{q\omega},
\end{eqnarray}
where ${\bf G}^K= 2{\bf G}^< + {\bf G}^R-{\bf G}^A$ is the Keldysh Green function, and
${\bf M}^q\equiv M^q_{\alpha\beta}$.

In our model the retarded vibron function is calculated from
the Dyson-Keldysh equation
\begin{equation}
 D^R(q,\omega)=\frac{2\omega_q}{\omega^2-\omega^2_q-2\omega_q{\Pi}^R(q,\omega)},
\end{equation}
where $\Pi(q,\omega)$ is the polarization operator (boson self-energy). The equation
for the lesser function (quantum kinetic equation in the integral form) is
\begin{equation}\label{kinPi}
 (\Pi^{R}_{q\omega}-\Pi^A_{q\omega})D^<_{q\omega}-(D^{R}_{q\omega}-D^A_{q\omega})
 \Pi^{<}_{q\omega}=0,
\end{equation}
this equation in the stationary case considered here is algebraic in the frequency domain.

The polarization operator is the sum of two parts, environmental and electronic:
$\Pi^{R,<}_{q\omega}=\Pi^{R,<(\rm env)}_{q\omega}+\Pi^{R,<(\rm el)}_{q\omega}$.

The environmental equilibrium part of the polarization
operator can be approximated by the simple expressions
\begin{eqnarray}
& \displaystyle \Pi^{R(\rm env)}(q,\omega)=-\frac{i}{2}\gamma_q{\rm sign}(\omega),
\\[0.3cm] & \displaystyle \Pi^{<(\rm env)}(q,\omega)=-i\gamma_qf_B^0(\omega){\rm
sign}(\omega),
\end{eqnarray}
where $\gamma_g$ is the vibronic dissipation rate, and $f_B^0(\omega)$ is the
equilibrium Bose-Einstein distribution function.

The electronic contribution to the polarization operator
within the SCBA is
\begin{align}\label{PiRA} \Pi^{R(\rm el)}(q,\omega)=
-i\int\frac{d\epsilon}{2\pi}{\rm Tr} & \left({\bf M}^q{\bf
G}^{<}_{\epsilon}{\bf M}^q{\bf G}^{A}_{\epsilon-\omega}+
\right. \nonumber \\ & \left. +{\bf M}^q{\bf
G}^R_{\epsilon}{\bf M}^q{\bf
G}^{<}_{\epsilon-\omega}\right),\\[0.3cm]
\Pi^{<(\rm el)}(q,\omega)= -i\int\frac{d\epsilon}{2\pi}{\rm Tr} &
\left({\bf M}^q {\bf G}^<_{\epsilon}{\bf M}^q{\bf
G}^>_{\epsilon-\omega}\right). \end{align}

We obtained the full set of equations, which can be used for numerical calculations.
We simplify these equations and obtain some analytical results in the {\em vibronic
quasiparticle approximation}, which assumes weak electron-vibron coupling limit and
weak external dissipation of vibrons:
\begin{equation}
 \gamma^*_q=\gamma_q-2{\rm Im}\Pi^{R}(\omega_q)\ll\omega_q.
\end{equation}
So that the spectral function of vibrons can be approximated by the Dirac $\delta$,
and the lesser function reads
\begin{equation} \label{qpv<}
 D^<(q,\omega)=-2\pi
 i\left[(N_q+1)\delta(\omega+\omega_q)+
 N_q\delta(\omega-\omega_q)\right],
\end{equation}
where $N_q$ is (nonequilibrium) number of vibrations in the $q$-th mode. So, in this
approximation the spectrum mo\-di\-fi\-ca\-tion of vibrons is not taken into account,
but the possible excitation of vibrations is described by the nonequilibrium $N_q$.
The dissipation of vibrons is neglected in the spectral function, but is taken into
account later in the kinetic equation for $N_q$. A similar approach to the
single-level problem was considered recently in \cite{MTU,GRN,Mitra04prb}. The more
general case with broadened equilibrium vibron spectral function seems to be not very
interesting, because in this case vibrons are not excited. Nevertheless, in the
numerical calculation it can be easy taken into consideration.

From the general quantum kinetic equation for vibrons (\ref{kinPi}) we obtain in this
limit
\begin{equation}
 N_q=\frac{\gamma_qN^0_q-{\rm Im}\Pi^{<}(\omega_q)}
 {\gamma_q-2{\rm Im}\Pi^{R}(\omega_q)}.
\end{equation}

This expression describes the number of vibrons $N_q$ in a nonequilibrium state,
$N^0_q=f_B^0(\omega_q)$ is the equilibrium number of vibrons. In the linear
approximation the polarization operator is independent of $N_q$ and $-2{\rm
Im}\Pi^{R}(\omega_q)$ describes additional dissipation. Note that in equilibrium
$N_q\equiv N^0_q$ because ${\rm Im}\Pi^{<}(\omega_q)=2{\rm Im}\Pi^{R}(\omega_q)
f_B^0(\omega_q)$. See also detailed discussion of vibron emission and absorption rates
in Refs.\,\cite{MTU}.

For weak electron-vibron coupling the number of vibrons is close to equilibrium and is
changed because of {\em vibron emission} by nonequilibrium electrons, $N_q$ is roughly
proportional to the number of such electrons, and the distribution function of
nonequilibrium electrons is not change essentially by the interaction with vibrons
(perturbation theory can be used). The situation changes, however, if nonequilibrium
dissipation $-2{\rm Im}\Pi^{R}(\omega_q)$ is {\em negative}. In this case the number
of vibrons can be essentially larger than in the equilibrium case ({\em vibronic
instability}), and the change of electron distribution function should be taken into
account self-consistently.

In the stationary state the {\em nonlinear} dissipation rate
\begin{equation}\label{negdiss}
 \gamma^*_q=\gamma_q-2{\rm Im}\Pi^{R}(\omega_q)
\end{equation}
is positive, but the nonequilibrium contribution to dissipation $-2{\rm
Im}\Pi^{R}(\omega_q)$ remains negative.

Additionally to the vibronic quasiparticle approximation, the {\em electronic
quasiparticle approximation} can be used when the coupling to the leads is weak. In
this case the lesser function can be parameterized through the number of electrons
$F_\eta$ {\em in the eigenstates of the noninteracting molecular Hamiltonian
$H^{(0)}_{M}$}
\begin{equation}\label{Gres}
 G^<_{\alpha\beta}=i\sum_{\gamma\eta}A_{\alpha\gamma}S_{\gamma\eta}F_\eta S^{-1}_{\eta\beta},
\end{equation}
we introduce the unitary matrix $\bf S$, which transfer the Hamiltonian ${\bf H}\equiv
H^{(0)}_{M\alpha\beta}$ into the diagonal form ${\bf\tilde H}={\bf S}^{-1}{\bf H}{\bf
S}$, so that the spectral function of this diagonal Hamiltonian is
\begin{equation}
 \tilde A_{\delta\eta}(\epsilon)=2\pi\delta(\epsilon-\tilde\epsilon_\delta)
 \delta_{\delta\eta},
\end{equation}
where $\tilde\epsilon_\delta$ are the eigenenergies.

\begin{figure}
\begin{center}
\epsfxsize=0.7\hsize
\epsfbox{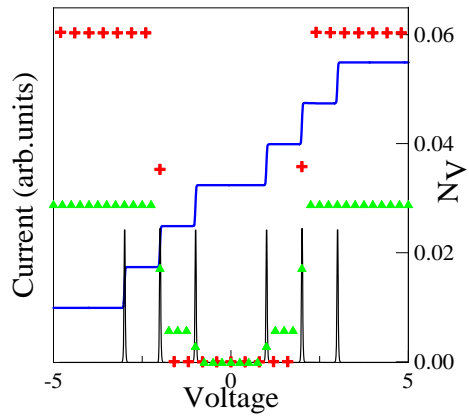}
\caption{(Color online) Vibronic emission in
the symmetric multilevel model: voltage-current curve, differential conductance, and
the number of excited vibrons in the off-resonant (triangles) and resonant (crosses)
cases (details see in the text).}
\label{fig2}
\end{center}
\end{figure}

Note that in the calculation of the self-energies and polarization operators we can
not use $\delta$-approximation for the spectral function (this is too rough and
results in the absence of interaction out of the exact electron-vibron resonance). So
that in the calculation we use actually (\ref{Gres}) with broadened equilibrium
spectral function. This approximation can be systematically improved by including
nonequilibrium corrections to the spectral function, which are important near the
resonance. It is important to comment that for stronger electron-vibron coupling {\em
vibronic side-bands} are observed in the spectral function and voltage-current curves
at energies $\tilde\epsilon_\delta\pm n\omega_q$, we do not consider these effects in
the rest of our paper and concentrate on resonance effects.

After correspondingly calculations we obtain finally
\begin{equation}\label{Nq}
 N_q=\frac{\gamma_qN^0_q-\sum_{\eta\delta}\kappa_{\eta\delta}(\omega_q)F_\eta(F_\delta-1)}
 {\gamma_q-\sum_{\eta\delta}\kappa_{\eta\delta}(\omega_q)(F_\eta-F_\delta)},
\end{equation}
where coefficients $\kappa_{\eta\delta}$ are determined by the spectral function and
electron-vibron coupling in the diagonal representation
\begin{equation}\label{Kappa}
 \kappa_{\eta\delta}(\omega_q)=\int\frac{d\epsilon}{2\pi}\tilde M^q_{\eta\delta}
 \tilde A_{\delta\delta}(\epsilon-\omega_q)\tilde M^q_{\delta\eta}\tilde
 A_{\eta\eta}(\epsilon),
\end{equation}
\begin{equation}
 F_\eta=\frac{\tilde\Gamma_{L\eta\eta}f^0_{L\eta}\!+\!\tilde\Gamma_{R\eta\eta}f^0_{R\eta}\!+\!
 \sum_{q\eta}\left[\zeta^{-q}_{\eta\delta}F_\delta N_q\!+\!
 \zeta^{+q}_{\eta\delta}F_\delta(1\!+\!N_q)\right]}
 {\tilde\Gamma_{L\eta\eta}\!+\!\tilde\Gamma_{R\eta\eta}\!+\!
 \sum_{q\eta}\left[\zeta^{-q}_{\eta\delta}(1\!-\!F_\delta\!+\!N_q)\!+\!
 \zeta^{+q}_{\eta\delta}(F_\delta\!+\!N_q)\right]},
\end{equation}
\begin{equation}
 \zeta^{\pm q}_{\eta\delta}=\tilde M^q_{\eta\delta}\tilde
 A_{\delta\delta}(\tilde\epsilon_\eta\pm\omega_q)
 \tilde M^q_{\delta\eta},
\end{equation}
here $\tilde\Gamma_{i\eta\eta}$ and $f^0_{i\eta}$ are the level width matrix in the
diagonal representation and Fermi function at energy $\tilde\epsilon_\eta-e\varphi_i$.

These kinetic equations are similar to the usual golden rule equations, but are more
general.

Now let us consider several examples of vibron emission and vibronic instability.

\begin{figure}
\begin{center}
\epsfxsize=0.7\hsize
\epsfbox{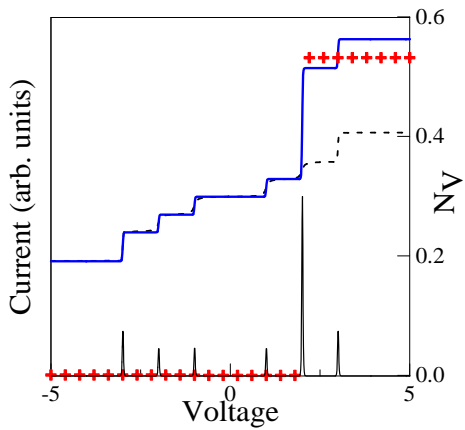}
\caption{(Color online) Vibronic instability
in an asymmetric multilevel model: voltage-current curve, differential conductance,
and the number of excited vibrons (crosses). Dashed line show the voltage-current
curve without vibrons (details see in the text).}
\label{fig3}
\end{center}
\end{figure}

\paragraph{(i) Vibron emission.}

First we consider the most simple case, when the instability is not possible and only
vibron emission takes place. This corresponds to a negative imaginary part of the
electronic polarization operator: ${\rm Im}\Pi^{R)}(\omega_q)<0$. From the Eq.
(\ref{Kappa}) one can see that for any two levels with the energies
$\tilde\epsilon_\eta>\tilde\epsilon_\delta$ the coefficient $\kappa_{\eta\delta}$ is
larger than $\kappa_{\delta\eta}$, because the spectral function $\tilde
A_{\delta\delta}(\epsilon)$ has a maximum at $\epsilon=\tilde\epsilon_\delta$. The
contribution of $\kappa_{\eta\delta}(\omega_q)(F_\eta-F_\delta)$ is negative if
$F_\eta<F_\delta$. This takes place in equilibrium, and in nonequilibrium for
transport through {\em symmetric} molecules, when higher energy levels are populated
after lower levels. The example of such a system is shown in Fig.\,\ref{fig2}. Here we
consider a simple three-level system ($\tilde\epsilon_1=1$, $\tilde\epsilon_2=2$,
$\tilde\epsilon_3=3$) coupled symmetrically to the leads
($\Gamma_{L\eta}=\Gamma_{R\eta}=0.01$). The current-voltage curve is the same with and
without vibrations in the case of symmetrical coupling to the leads and in the weak
electron-vibron coupling limit (if we neglect change of the spectral function). The
figure shows how vibrons are excited, the number of vibrons $N_V$ in the mode with
frequency $\omega_0$ is presented in two cases. In the off-resonant case (green
triangles) $N_V$ is very small comparing with the resonant case
($\omega_0=\tilde\epsilon_2-\tilde\epsilon_1$, red crosses, the vertical scale is
changed for the off-resonant points). In fact, if the number of vibrons is very large,
the spectral function and voltage-current curve are changed. We shall consider this in
a separate publication.

\paragraph{(ii) Vibronic instability.}

Now let us consider the situation when the imaginary part of the electronic
polarization operator can be positive: ${\rm Im}\Pi^{R}(\omega_q)>0$. Above we
considered the normal case when the population of higher energy levels is smaller than
lower levels. The opposite case $F_2>F_1$ is known as inversion in laser physics. Such
a state is unstable if the total dissipation $\gamma^*_q$ (\ref{negdiss}) is negative,
which is possible only in the nonstationary case. As a result of the instability, a
large number of vibrons is excited, and in the stationary state $\gamma^*_q$ is
positive. This effect can be observed for transport through {\em asymmetric}
molecules, when higher energy levels are populated {\em before} lower levels. The
example of a such system is shown in Fig.\,\ref{fig3}. It is the same three-level
system as before, but the first and second levels are coupled not symmetrically to the
leads ($\Gamma_{L1}=0.001$, $\Gamma_{R1}=0.1$, $\Gamma_{L2}=0.1$,
$\Gamma_{R2}=0.001$). The vibron couple resonantly these levels
($\omega_q=\tilde\epsilon_2-\tilde\epsilon_1$). The result is qualitatively different
from the symmetrical case. The voltage-current curve is now asymmetric, a large {\em
step} corresponds to the resonant level with inverted population.

Note the importance of the off-diagonal electron-vibron coupling for the resonant
effects. If the matrix $\bf\tilde M$ in the eigen-state representation is diagonal,
there is no resonant coupling between different electronic states.

\begin{figure}
\begin{center}
\epsfxsize=0.7\hsize
\epsfbox{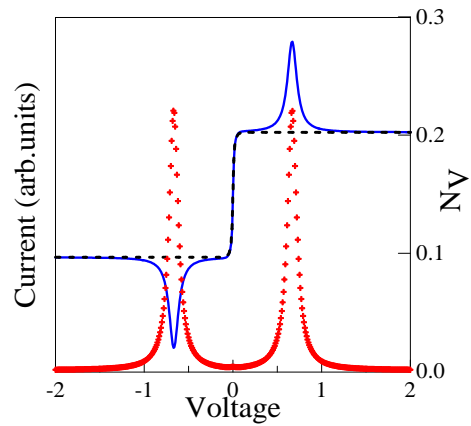}
\caption{(Color online) Floating level
resonance: voltage-current curve and the number of excited vibrons (crosses). Dashed
line show the voltage-current curve without vibrons (details see in the text).}
\label{fig4}
\end{center}
\end{figure}

\paragraph{(iii) Floating-level resonance.}

Finally, let us consider the important case, when initially symmetric molecule becomes
asymmetric when the external voltage is applied. The reason for such asymmetry is
simply that in the external electric field left and right atoms feel different
electrical potentials and the position of the levels
$\epsilon_\alpha=\epsilon^{(0)}_\alpha+e\varphi_\alpha$ is changed (float) with the
external voltage. The example of a such system is shown in Fig.\,\ref{fig4}. Here we
consider a two-level system, one level is coupled electrostatically to the left lead
$\tilde\epsilon_1\propto \varphi_L$, the other level to the right lead
$\tilde\epsilon_2\propto \varphi_R$, the tunneling coupling to the leads also is not
symmetrical ($\Gamma_{L1}=0.1$, $\Gamma_{R1}=0.001$, $\Gamma_{L2}=0.001$,
$\Gamma_{R2}=0.1$). The frequency of the vibration, coupling these two states, is
$\omega_0=1$. When we sweep the voltage, a {\em peak} in the voltage-current curve is
observed when the energy difference $\tilde\epsilon_1-\tilde\epsilon_2\propto eV$ is
going through the resonance $\tilde\epsilon_1-\tilde\epsilon_2\approx\omega_0$.

In conclusion, we considered the excitations of quantum molecular vibrations in the
nonequilibrium state and their influence on the voltage-current curves of a single
molecule placed between two equilibrium leads. The importance of vibron emission and
vibronic instability in molecular transport is demonstrated.

We thank J. Keller and K. Richter for valuable discussions. This work was supported by
the Volkswagen Foundation under grant I/78~340 and by the EU under contract
IST-2001-38951.

\end{document}